# Magnetization controlled by crystallization in soft magnetic Fe-Si-B-P-Cu alloys

Hiroshi Nakajima[1], Akihiro Osako[1], Noriharu Yodoshi[2], Yoshiharu Yamada[3], Hirofumi Tsukasaki[1], Ken Harada[4], Yuki Sakai[5, 6], Kei Shigematsu[5, 6], Takumi Nishikubo[5, 6], Masaki Azuma[5, 6], and Shigeo Mori[1]

[1]*Department of Materials Science, Osaka Metropolitan University, Sakai, Osaka 599-8531, Japan*

[2]*Department of Mechanical Engineering, Kyushu University, Fukuoka, Japan*

[3]*Osaka Research Institute of Industrial Science and Technology, Izumi, Osaka, 594-1157 Japan*

[4]*Center for Emergent Matter Science (CEMS), Institute of Physical and Chemical Research (RIKEN), Hatoyama, Saitama 350-0395, Japan*

[5]*Laboratory for Materials and Structures, Tokyo Institute of Technology, Nagatsuta 4259, Midori-ku, Yokohama 226-8503, Japan*

[6]*Kanagawa Institute of Industrial Science and Technology, Shimoimaizumi 705-1, Ebina, Kanagawa 243-0435, Japan*

ABSTRACT

Soft magnetic materials have low coercive fields and high permeability. Recently, nanocrystalline alloys obtained using annealing amorphous alloys have attracted much interest since nanocrystalline alloys with small grain sizes of tens of nanometers exhibit low coercive fields comparable to that of amorphous alloys. Since nanocrystalline soft magnetic materials attain remarkable soft magnetic properties by controlling the grain size, the crystal grains' microstructure has a substantial influence on the soft magnetic properties. In this research, we examined the magnetic properties of Fe-Si-B-P-Cu nanocrystalline soft magnetic alloys obtained by annealing amorphous alloys. During crystallization, the observation findings reveal the correlation between the generated microstructures and soft magnetic properties.

# Introduction

Soft magnetic materials possess a low coercive field and high permeability [1]. Low core loss (*i.e.,* low coercive field) is preferable under alternate-current magnetic fields when a soft magnetic material is employed. Amorphous alloys are known for low coercive fields among soft magnetic materials. However, they have thermal instability since amorphous alloys are thermodynamically nonequilibrium states (metastable phases). Recently, nanocrystalline alloys generated through heat treatment of amorphous alloys have attracted attention to solving this problem. When the grain size is as small as 10 nm, the coercive field of nanocrystalline alloys reduces in proportion to the grain size's sixth power, and the values are comparable to those of amorphous alloys [2]. They also demonstrate higher permeability and higher saturation magnetic flux density than amorphous alloys. Nanocrystalline soft magnetic materials attain remarkable soft magnetic properties by regulating the crystal grain size and the crystal grains' microstructure has a substantial influence on the soft magnetic properties.

Amorphous soft magnetic alloys that are presently in practical use include Fe-Si-B amorphous alloys. However, their grains tend to coarsen upon annealing treatment. Earlier investigations have demonstrated that adding P or Cu to these alloys can fine the crystal grains [3,4]. Furthermore, the Fe-Si-B-P-Cu alloys exhibited remarkable soft magnetic properties as nanocrystalline soft magnetic materials [5–8]. Compared to silicon steel, the Fe-Si-B-P-Cu soft magnetic material has a high saturation magnetic flux density while maintaining low iron loss and high magnetic permeability [9]. It is crucial to clarify the relationship between the microstructure of nanocrystalline alloys and the magnetic properties in the Fe-Si-B-P-Cu alloys [10–12]. However, in terms of real-space observation of microstructures and magnetic domains, not enough investigation has been performed. In this research, we aimed to clarify the mechanisms that influence coercive fields in nanocrystalline soft magnetic $(Fe_{76}Si_9B_{10}P_5)_{99.5}Cu_{0.5}$ alloys, which is a composition exhibiting superior soft magnetic properties. We heat-treated the amorphous Fe-Si-B-P-Cu alloys and examined the relationship between microstructural modifications associated with crystallization and changes in magnetic properties.

# Experimental methods

To prepare $(Fe_{76}Si_9B_{10}P_5)_{99.5}Cu_{0.5}$, high-purity materials of Fe, Si, B, and Fe-P were weighed to obtain the specified ratio. In an Ar gas atmosphere, the material was alloyed using high-frequency melting. A single roll liquid quenching method was used to amorphize the alloyed specimen in the Ar gas atmosphere. Copper rolls were employed at a rotation speed of 4000 rpm. A quartz glass nozzle with a hole diameter of 0.5 mm was used for the quenching. By heating the amorphous specimen, annealed specimens were produced after being sealed in evacuated silica tubes. To achieve thermal equilibration, the target temperature of an electric furnace was maintained for more than two hours and then the specimen was inserted into the furnace and maintained for the desired annealing time. The specimen was quenched by eliminating it from the furnace and cooled at room temperature. Differential Scanning Calorimetry (DSC) measurements were conducted in an $N_2$ atmosphere to examine the specimen's crystallization temperature at a heating speed of 10 °C/min (Rigaku: Thermo plus EVO2 DSC8231). High-temperature synchrotron X-ray diffraction (XRD) was conducted up to 30 °C–700 °C by spraying $N_2$ at the wavelength $\lambda = 0.40990$ nm in the Beamline 02B2 of SPring-8 [13]. To obtain magnetic hysteresis loops, Vibrating Sample Magnetometer (VSM) measurements were conducted at room temperature using a VSM instrument (BHV-50 by Riken Denshi Co., Ltd.). The coercive field was assessed from the hysteresis loops. Transmission electron microscopy (TEM) was conducted using JEOL-2100F (JEOL Co. Ltd.). Bright-field imaging and electron diffraction were observed to show the microstructure. Magnetic domain walls were visualized using Fresnel imaging of Lorentz microscopy by defocusing an imaging lens. The applied magnetic fields were adjusted using the objective lens in the Fresnel mode [14]. Thin specimens were constructed by argon ion milling using a Precision Ion Polishing System (PIPS) manufactured by Gatan. The beam energy was set to 4 keV. After holes were produced in the specimen, the surfaces were treated by irradiating the beam at 0.5 keV for about 30 min to eliminate damaged layers.

**Results and discussion**

First, DSC and in situ XRD measurements were conducted to measure the crystallization temperature and identify precipitated phases. Figure 1(a) depicts DSC measurements of the $(Fe_{76}Si_9B_{10}P_5)_{99.5}Cu_{0.5}$ specimens. In the as-prepared amorphous specimen, two exothermic peaks were observed at 502 °C and 537 °C. DSC measurements were also conducted on annealed specimens. The annealing treatment temperature was set to 500 °C since the temperature can readily regulate the crystallization using annealing time and complete crystallization occurred for a very short time above 500 °C. The first exothermic peak remained in the specimen with a heat treatment time of 2 min, whereas the first peak disappeared in that with a heat treatment time of 4 min. Furthermore, after 5 min of heat treatment, the second peak's area became smaller in the specimens, demonstrating that crystallization had progressed. Additionally, in the specimen with a heat treatment time of 20 min, the peak disappeared, indicating that almost all areas were crystallized. However, another exothermic peak is more pronounced at 580 °C in the specimen annealed for 5–20 min. The DSC measurements demonstrate that the crystallinity can be regulated using the annealing time and temperature.

To detect the precipitated phases, high-temperature synchrotron XRD measurements were conducted on a $(Fe_{76}Si_9B_{10}P_5)_{99.5}Cu_{0.5}$ alloy. Figure 1(b) depicts in situ XRD profiles at elevated temperatures. The peak appeared when heated to 550 °C and the peak intensity increased with further heating, confirming that crystallization was in progress. The broad scattering observed from room temperature to 550 °C became a peak (110 reflection) at 550 °C. The peaks at 550 °C and 600 °C can be assigned to α-Fe precipitates (the space group $Im\bar{3}m$) and $Fe_3B$ precipitates (the space group $Pnma$), respectively. Comparing the DSC and XRD findings, the XRD profiles show that the peaks at 502 °C and 537 °C match the α-Fe's precipitation. The broad peak at 580°C may be ascribed to the precipitation of $Fe_3B$. Thus, DSC measurements and high-temperature synchrotron XRD findings show that the $(Fe_{76}Si_9B_{10}P_5)_{99.5}Cu_{0.5}$ alloy undergoes two-step crystallization upon heat treatment. The crystallization's first stage and second stage occurred through the precipitation of α-Fe and $Fe_3B$, respectively.

Further investigations were performed with isothermal-annealed specimens since changing the annealing time at 500 °C can regulate the crystallization systematically as shown in the DSC measurements. VSM measurements were conducted to examine the magnetic properties of $(Fe_{76}Si_9B_{10}P_5)_{99.5}Cu_{0.5}$ alloy annealed at 500°C. Figure 1(c) shows that the coercive field can be modified upon the annealing treatments. The coercive field's magnitude can be generated from the *x* axis' intersection in magnetization. Figure 1(d) depicts the change in coercive fields by annealing. The coercive field begins to increase when the annealing treatment time exceeds 3 min and increases rapidly after 6 min. The specimen with an annealing treatment time of 20 to 1000 min demonstrated a high coercive field of approximately 13000 A/m. The coercive field tended to reduce again after 1000 min. No further modification in the coercive field was observed after 12000 min of annealing treatment.

TEM observations were conducted in $(Fe_{76}Si_9B_{10}P_5)_{99.5}Cu_{0.5}$ alloys with annealing times of 4 to 20 min, which demonstrated an increase in the coercive field. Figure 2 depicts bright-field images and electron diffraction patterns for specimens with annealing for 4, 12, and 20 min. The electron diffraction pattern indicates that the specimen of 4 min comprised the amorphous phase and α-Fe phase. In the specimens after 12 min of annealing treatments, diffraction spots of $Fe_3B$ were observed in the electron diffraction patterns (red circles) although the diffraction intensity was weak. The findings propose that the specimens with heat treatment time after 12 min were a mixture of α-Fe and $Fe_3B$ phases. No halo-ring pattern was observed in the electron diffraction pattern for the specimen with a heat treatment time of 20 min, showing that the sample was a crystal without an amorphous region. However, no substantial difference in the average grain size can be observed among the samples with heat treatment times of 4, 12, and 20 min. Although Figure 1(d) depicts that the coercive field increased drastically after 6 min of heat treatment time, the average grain was increased only by 4.6 nm, as shown in the observation results. The findings indicate that the ratio of crystalline and amorphous areas may influence the coercive field rather than the grain size. Furthermore, the TEM observations indicated the precipitation of $Fe_3B$ after 5 min of annealing treatment, agreeing with the DSC and XRD measurements. Therefore, $Fe_3B$ grains could also contribute to the increase in the coercive field.

We examined the relationship between the grains and magnetic-wall movement under applied magnetic fields. In the specimen with an annealing treatment time of 12 min, magnetic domain walls were observed to be pinned at grain boundaries. Figure 3(a)–(e) depicts Fresnel images at the increasing magnetic fields. The bright lines denote magnetic domain walls (yellow arrowheads). The walls were primarily formed in the amorphous phase and grain boundaries. The magnetic domain walls were moved by increasing the magnetic fields. However, one of the walls was pinned as illustrated in the red circle point. Figure 3(f) depicts a bright-field image around the grains at the pinned magnetic wall. The magnetic domain wall was rotated at a grain with an increasing magnetic field. The dark-field imaging using a $Fe_3B$ reflection (the inset) visualized the grain as bright, demonstrating that the grain is $Fe_3B$. Therefore, our microscopy observations confirmed that $Fe_3B$ precipitates are one of the causes of the increased coercive field. The magnetic anisotropy of $Fe_3B$ varies from that of α-Fe, and consequently, at $Fe_3B$ precipitates magnetic domain walls are likely to be pinned [15]. The pinning by $Fe_3B$ may explain the rapid increase in coercive fields of the specimen after 6 min of annealing treatment.

Additionally, the relationships between the grains and magnetic domains were investigated for annealing treatment times of 600 and 4200 min, which indicated high coercive fields of approximately 13,000 A/m although the specimen of 4200 min demonstrated a decreased coercive field. Figure 4(a) and 4(d) depict bright-field images and electron diffraction patterns. Comparing the specimens with an annealing treatment time of 20 min (Figure 2), the specimens with a heat treatment time of 600 and 4200 min had larger average grain sizes of 33.5 nm and 44.3 nm, respectively. These values are comparable to the grain sizes of similar compositions [4]. No halo-ring pattern was observed in the electron diffraction patterns in both the specimens, showing that the specimens comprised α-Fe and $Fe_3B$ phases.

The coercive field of the specimen annealed for 4200 min (9760 A/m) was smaller than that of the specimen annealed for 600 min (13306 A/m). Magnetic domain walls were observed to show the reason [Fig. 4(b) and 4(e)]. The schematic of Fig. 4(c) illustrates the Fresnel image of the 600-min specimen and indicates that magnetic domain walls existed at the grain boundaries. Consequently, the magnetic domain walls should be challenging to move by magnetic fields since the grain boundaries pin the walls, which will increase the coercive field. The 4200-min-annealed specimen's Fresnel image demonstrates that magnetic walls were primarily formed at grain boundaries, whereas some magnetic walls were formed within grains, as depicted by the red lines in Fig. 4(f). The coercive field of the specimen of 4200 min has likely decreased since no pinning influence occurred for the magnetic domain walls that existed within the grains. The bright-field images suggest that the threshold of the grain size that this phenomenon occurs lies between 33.5 nm and 44.3 nm in this material. These findings reveal that when the grain size exceeds a certain size, magnetic domain walls are formed in grains, and as a result, the coercive field decreases since the magnetic walls are more likely to move. This explains the change in the coercive field in Fig. 1(d).

## Conclusions

To clarify the mechanism of coercive field change in Fe-Si-B-P-Cu nanocrystalline soft magnetic alloys, we observed the microstructure of heat-treated $(Fe_{76}Si_9B_{10}P_5)_{99.5}Cu_{0.5}$ alloy and examined the relationship between the microstructure and magnetic properties. The coercive field of $(Fe_{76}Si_9B_{10}P_5)_{99.5}Cu_{0.5}$ amorphous alloy increased with annealing treatment longer than 6 min and tended to reduce as the grain size increased to a certain degree. TEM observation showed the mechanisms that contribute to the increase in coercive fields. One effect is the pinning of magnetic domain walls by $Fe_3B$ precipitates. The specimen with decreased coercive fields demonstrated magnetic walls formed within crystal grains. This magnetic configuration has no barrier to the magnetic walls' movement within the grains, resulting in a smaller coercive field.

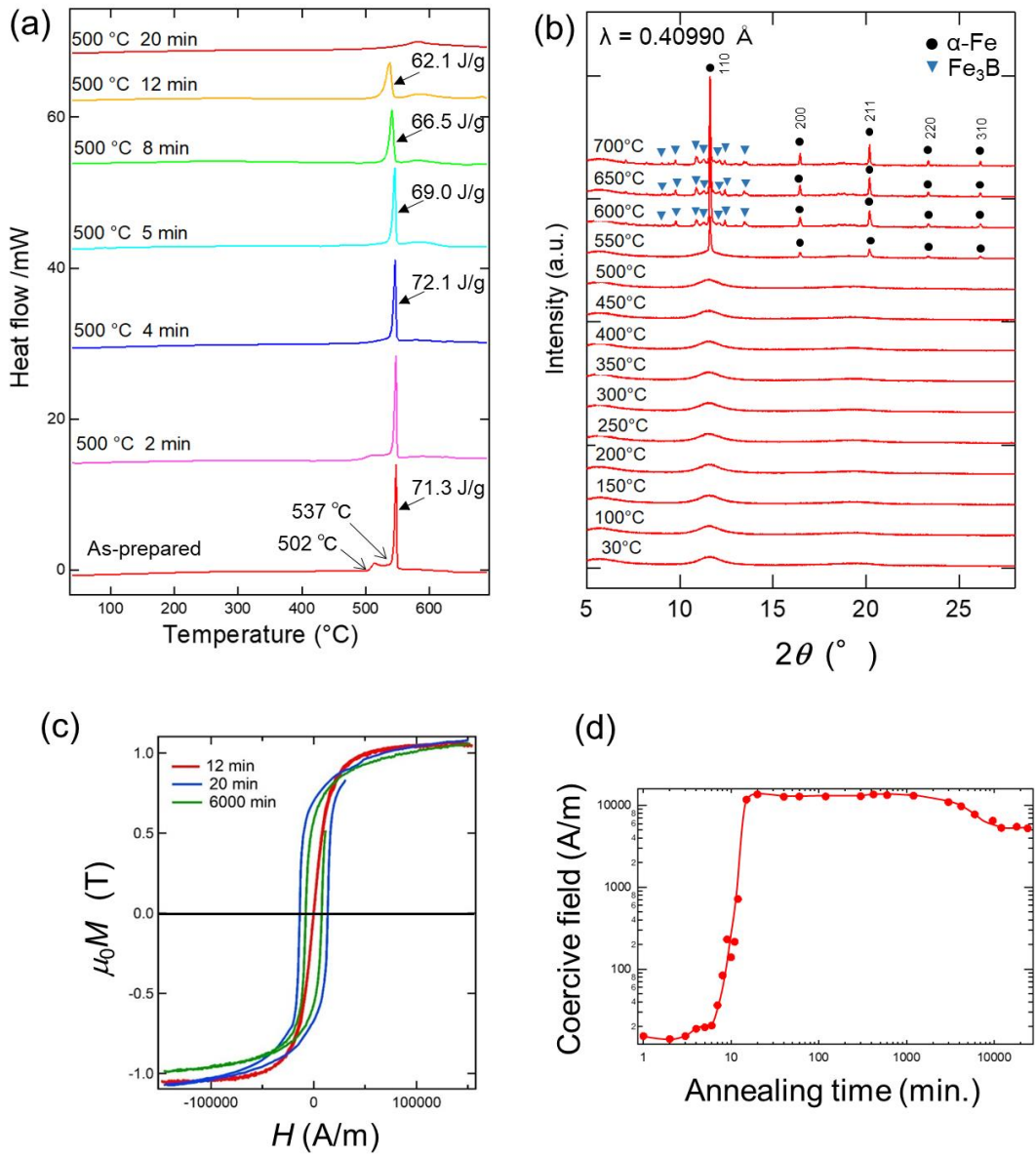

Figure 1. (a) DSC measurements of as-prepared and annealed specimens. (b) in situ heating synchrotron X-ray diffraction (c) The magnetic field ($H$) dependence of the magnetization ($\mu_0 M$) in annealed specimens. (d) The annealing time dependence of coercive fields.

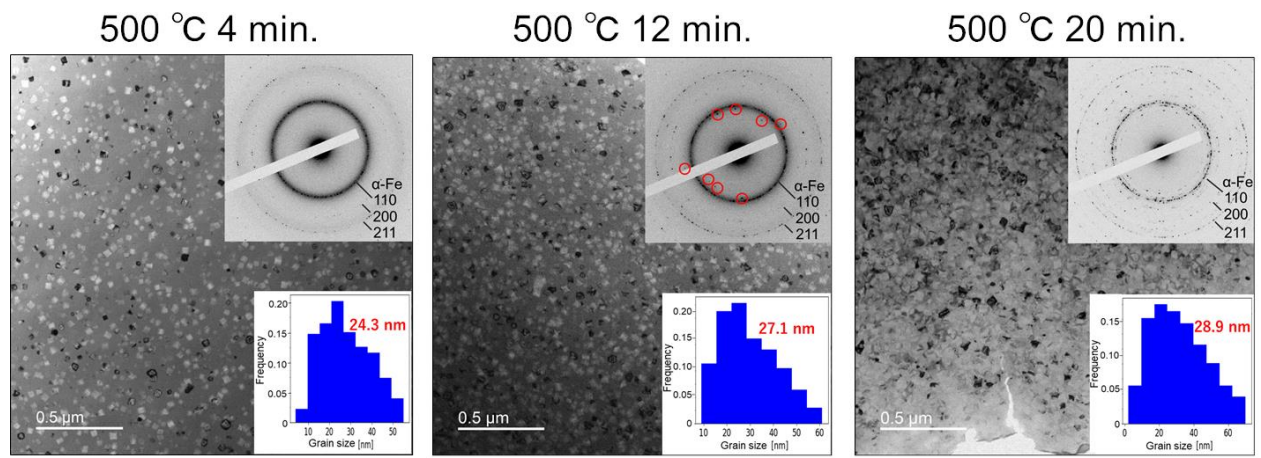

Figure 2. Bright-field images of annealed specimens. The insets show electron diffraction patterns and grain size distributions with the average size listed. The red circles represent Bragg reflections of $Fe_3B$.

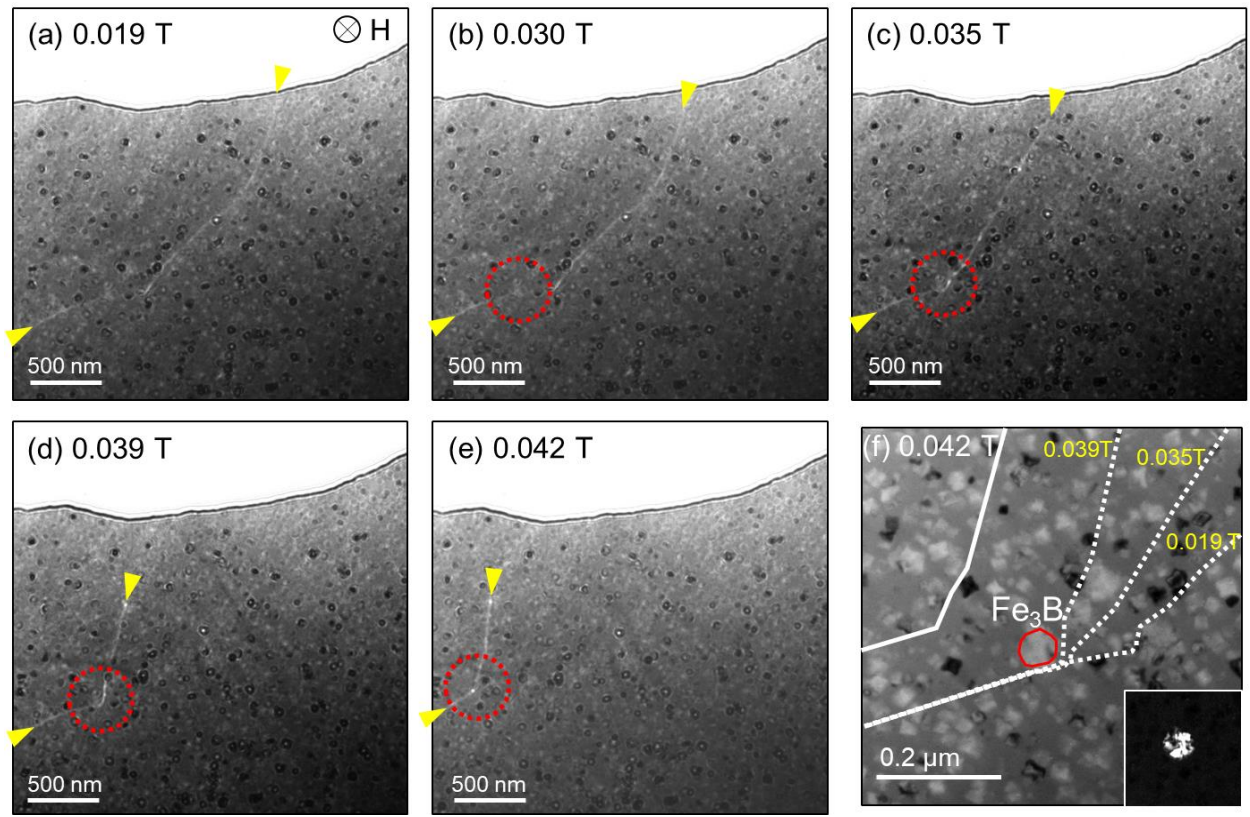

Figure 3. (a)–(e) Applied magnetic field dependence of magnetic domain walls in the specimen annealed for 12 minutes. The dotted circle is the place where the magnetic domain wall is pinned. (f) Bright-field image around the red circled region. The dotted lines represent the trace of the magnetic domain walls at 0.019–0.039T. The solid line indicates the magnetic domain wall position at 0.042T. The inset is a dark-field image using an $Fe_3B$ reflection around the marked region.

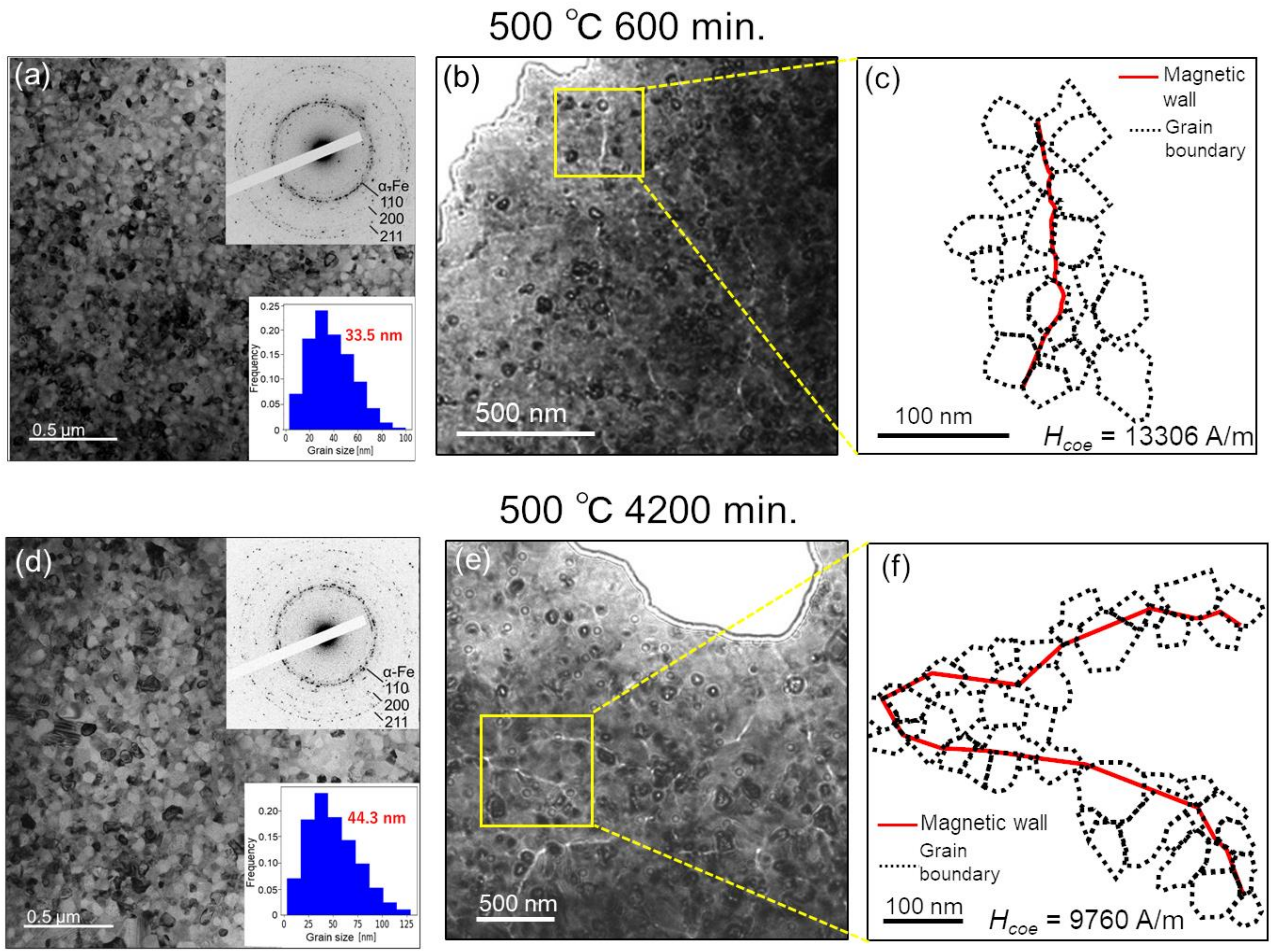

Figure 4. Observation results of long-time annealed specimens (600 and 4200 min). (a, d) bright-field images. The insets are electron diffraction patterns and grain size distributions with the average size. (b, e) Fresnel images visualizing magnetic domain walls. (c, f) Schematics of the yellow-marked regions.